\documentclass[twocolumn]{aastex6}

\begin{document}

\title{Is the GW150914-GBM really associated with the GW150914?}

\author{Shaolin Xiong}
\affil{Key Laboratory of Particle Astrophysics, Institute of High Energy Physics, Chinese Academy of Sciences, Beijing, China}
\email{xiongsl@ihep.ac.cn}

\begin{abstract}
Finding the electromagnetic (EM) counterpart is critically important for a gravitational wave event. Although many efforts have been made to search for the purported EM counterpart of GW150914, the first gravitational wave event detected by LIGO, only Fermi/GBM reported an excess above background (i.e. GW150914-GBM) at 0.4 s after the LIGO trigger time, that is possibly associated with this GW event \citep{connaughton_fermi_2016}. However, since there is no significant detection by the INTEGRAL/SPI-ACS around the time of GW150914-GBM, a great debate has been raised about whether GW150914-GBM is of astrophysical origin and associated with the GW150914 \citep{savchenko_integral_2016}. In order to answer this question, we re-analyzed the GBM data with a straightforward but sophisticated method. We find that the excess of GW150914-GBM mostly comes from those detectors with bad viewing angles to the GW event, whereas the good viewing detectors see nothing significant beyond background fluctuation around the trigger time of GW150914. This finding suggests that GW150914-GBM is very unlikely associated with the GW150914. Given that GW150914-GBM is the only event found by GBM that is possibly associated with this GW event in a comprehensive search, we conclude that GBM did not detect any electromagnetic radiation from the GW150914.

\end{abstract}


\section{Introduction} \label{sec:intro}

The era of gravitational wave (GW) astronomy has come since the Laser Interferometer Gravitational-wave Observatory (LIGO) detected the first GW event, GW150914, which is thought to be produced by the merging of a stellar mass binary black hole system \citep{the_ligo_scientific_collaboration_observation_2016}. Finding the electromagnetic (EM) counterpart of the GW event is critically important in many aspects, such as identification of the GW event, study of the progenitor and its environment, fundamental physics and cosmology.

Although many efforts have been made to search for the purported EM counterpart of GW150914 \citep{abbott_localization_2016}, only the Gamma-ray Burst Monitor (GBM) onboard Fermi \citep{meegan_fermi_2009} recorded an excess (denoted as GW150914-GBM) at 0.4 s after the LIGO trigger time, which is claimed to be possibly associated with this GW event  \citep{connaughton_fermi_2016}. However, since there is no significant signal detected by the INTEGRAL/SPI-ACS around the time of GW150914-GBM \citep{savchenko_integral_2016}, a great debate has been raised about whether GW150914-GBM is of astrophysical origin and associated with the GW150914.

\citet{connaughton_fermi_2016} reported that GW150914-GBM is likely associated with the GW150914 based on the following reasons: (a) Its time is only 0.4 s after the LIGO trigger time of the GW event. (b) Its localization is not inconsistent with LIGO locations of GW150914, although quite poor due to the weakness and unfavorable incident angles to GBM detectors. (c) Its duration and spectrum are consistent with typical short Gamma-ray Bursts (GRBs) which is thought to be candidate of GW EM counterpart. (d) It cannot be attributed to other known sources, such as solar activity, terrestrial gamma-ray flashes \citep{briggs_terrestrial_2013} or terrestrial electron beams \citep{xiong_location_2012} produced on the Earth, magnetosphereic activities and galactic sources.

However, \citet{savchenko_integral_2016} argued that the non-detection of SPI-ACS suggests the GW150914-GBM is likely from magnetospheric activity or background fluctuation. But \citet{connaughton_fermi_2016} argued that the fluence of GW150914-GBM is consistent with the upper limit estimated by SPI-ACS for most regions of LIGO arc; about half of weak GBM events are not found by SPI-ACS; the origin of magnetosphere is very unlikely because GBM is on the low magnetic latitude and the duration of GW150914-GBM is too short for the magnetospheric events usually seen by GBM. However, the reported GBM false alarm rate of 0.0022 \citep{connaughton_fermi_2016} is only of moderate significance to reject the possibility of background variation.

Given that there is no other measurement in a similar energy range as GBM and SPI-ACS to the GW150914 region around the LIGO trigger time, and that no other EM counterpart detection has been reported yet, the origin of GW150914-GBM and whether GW150914 has an EM counterpart remain a big mystery, but of high importance. Here we focus on the observational properties of GW150914-GBM by carefully analyzing the GBM data using a straightforward but sophisticated method.

\section{Re-analysis of the GBM data} \label{sec:reanalysis}

GBM employs the Thallium-doped Sodium Iodide (NaI) and Bismuth Germanate (BGO) scintillation detectors to cover a wide energy range from 8 keV to 40 MeV \citep{meegan_fermi_2009}. Since a NaI detector only has a good viewing to sources within $\sim60\arcdeg$ of the detector's normal, the response of the detector decreases rapidly when the incident angle is greater than 60$\arcdeg$ \citep{meegan_fermi_2009, connaughton_fermi_2016}. To form an all-sky field of view (FOV) except the inevitable blocking by the Earth in Low Earth Orbit, GBM consists of 12 NaI detectors with different orientations placed at the four corners of the spacecraft and two BGO detectors set in the opposite sides of the spacecraft \citep[see][fig. 4]{meegan_fermi_2009}. Such configuration is also used to localize astrophysical transients based on the fact that these detectors will receive different number of photons from a distant source due to different incident angles of the source to different detectors. Such localization method \citep{connaughton_localization_2015} has successfully helped GBM to localize $\sim$ 2000 GRBs \citep{kienlin_second_2014}.

\begin{figure}[ht!]
\figurenum{1}
\plotone{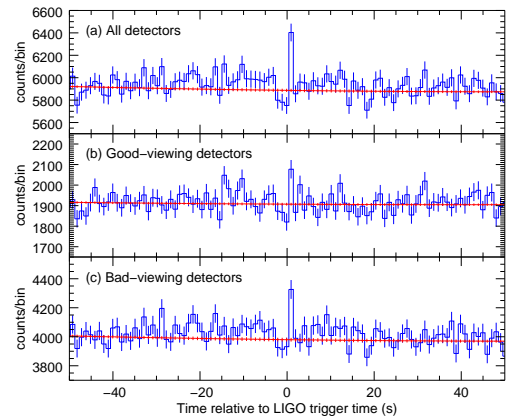}
\caption{Light curve of the GW150914-GBM at T0+0.4 s. (a) summed light curve for all 14 detectors.
(b) summed light curve for good-viewing detectors (NaI 2, NaI 4, NaI 5, BGO 0). (c) summed light curve for bad-viewing detectors (the other 10 detectors). Red lines are fitted background. \label{fig:f1}}
\end{figure}

The GBM team has reported an excess, GW150914-GBM, with false alarm rate of 0.0022 at T0+0.4 s, where T0 is the LIGO trigger time 2015-09-14 09:50:45.391 UTC, and this event is believed to be possibly linked to GW150914 \citep{connaughton_fermi_2016}. As shown in the top panel in Figure \ref{fig:f1}, we produced the summed light curve of all 14 detectors for GW150914-GBM in the same energy selection as used in \citet{connaughton_fermi_2016}, which is very consistent with the results (bottom panel of Fig. 4) in \citet{connaughton_fermi_2016}.

\begin{figure}[ht!]
\figurenum{2}
\plotone{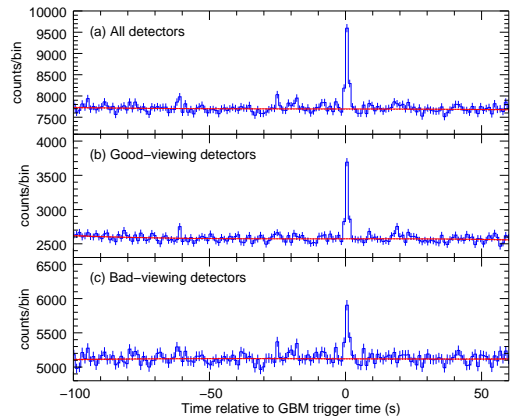}
\caption{Light curve of the GRB140716A. Other captions are the same as Figure \ref{fig:f1}.  \label{fig:f2}}
\end{figure}

Next we test whether GW150914-GBM is of astrophysical origin and associated with the GW event. Unlike the search algorithm utilized in \citet{connaughton_fermi_2016} which is based on a weighted light curve of all 14 detectors of GBM, here we choose to use a much more straightforward but sophisticated approach: we divide all 14 detectors into two groups with one group having good viewing angles to the LIGO location of the GW event and the other group bad viewing angles. According to the placement of the detectors described above, good-viewing detector group should have seen more photons than the bad-viewing group from the GW150914, if there is any high energy emission from it. We used the calculation of the incident angle for each detector in Table 2 of \citet{connaughton_fermi_2016} to decide which detector belongs to which group. It turns out that NaI 2, NaI 4, NaI 5 and BGO 0 have smaller viewing angles than other detectors for all possible locations of the GW event, thus these four detectors are chosen to be the good-viewing detectors for GW150914. The incident angle to the good-viewing group ranges from 20$\arcdeg$ to 80$\arcdeg$, depending on the real location of GW150914. We slightly relieved the usual threshold of good-viewing of 60$\arcdeg$ to 80$\arcdeg$, because there are too few detectors with less than 60$\arcdeg$ for a single location of the GW event. The other ten detectors compose the bad-viewing group because their incident angles are well greater than 80$\arcdeg$, resulting in either very little exposure to the GW event or completely blocked by the spacecraft.

Light curves for all 14 detectors, good-viewing and bad-viewing detectors are separately plotted in Figure \ref{fig:f1}. In the duration of GW150914-GBM (from T0+0.41 s to T0+1.43 s), there are 462 ($\sigma$ = 82.1, significance of this excess is 5.63 $\sigma$) net counts above the fitted background level in all 14 detectors, where 306 ($\sigma$ = 67.3, significance of this excess is 4.55 $\sigma$) of them come from bad-viewing detectors and 156 ($\sigma$ = 47.0, significance of this excess is 3.32 $\sigma$) from good-viewing ones. We suggest this event is very unlikely from the GW150914 as following:

\textbf{(I.) The majority of the excess ($\frac{306}{462}=66\%$) of GW150914-GBM is from bad-viewing detectors, which is inconsistent with an astrophysical source incident from the LIGO localizations visible to the GBM}.
The ratio of net counts of good-viewing and bad-viewing detectors for GW150914-GBM is 0.51 $\pm$ 0.19. In principle, one could do a detailed Monte Carlo simulation to estimate what this ratio should be for an astrophysical source; however, such simulation is impossible without the accurate mass model of the Fermi spacecraft which is unavailable for public. Alternatively, we estimate the expected ratio from GBM-Swift jointly-detected GRBs (shown in Table \ref{tab:grbs}) with similar incident angles in spacecraft coordinates as GW150914, because it only depends sensitively on the incident angle in the spacecraft coordinates for the spectra dominated by low-energy photons, which is always true for GRBs in GBM energy range. As an example, the light curve of GRB140716A is shown in Figure \ref{fig:f2}. The measured net counts ratio (0.51 $\pm$ 0.19) of GW150914-GBM is far away from that of GRBs in Table \ref{tab:grbs}, consistent with the fact that the best localization of GW150914-GBM is on the Earth, rather than the GW150914 locations \citep{connaughton_fermi_2016}. Assume the excess in bad-viewing detectors for GW150914-GBM is from LIGO locations visible to GBM, and we use 1.4 as a conservative estimation of the net counts ratio for the GW150914, then the expected net counts in good-viewing detectors should be 306 $\times$ 1.4 = 428.4 counts, with the error the same as original measurement of 67.3, because it is dominated by the background variation. Thus this assumption is rejected in $(428.4 - 156)/ \sqrt{67.3^2 + 47.0^2} = 3.3 ~\sigma$. 

Indeed, as shown in the Table 1 and 2 in \citet{connaughton_fermi_2016}, the bad-viewing detectors (NaI 9, BGO 1) with high incident angle for all possible LIGO locations have the most high significance of excess while the good-viewing detector (NaI 2) with much smaller incident angle has the least significance of excess. Such distribution of excess among detectors is inconsistent with an astrophysical source from LIGO locations. It's very likely that GW150914-GBM will not be found out by the targeted search \citep{connaughton_fermi_2016} without the excess signal in the bad-viewing detectors. 

\textbf{(II.) The excess of GW150914-GBM in good-viewing detectors is consistent with background fluctuation}.
To estimate the false alarm rate of the excess of GW150914-GBM in good-viewing detectors, we searched 216.6 ks of data of good-viewing detectors and found 56 events with higher significance, giving a false alarm rate of $2.6\times10^{-4}$ Hz, much higher than that reported in \citet{connaughton_fermi_2016}. Note that there are already several spikes with similar excess as GW150914-GBM within only 100 s of data in good-viewing detectors, see panel b in Figure \ref{fig:f1}. This suggests that the excess of GW150914-GBM in good-viewing detectors is consistent with background fluctuation.

In fact, the ratio of net counts between good-viewing and bad-viewing detectors for GW150914-GBM (0.51 $\pm$ 0.19) is approximately equal to the ratio of background level (red lines in panel b and c in Figure \ref{fig:f1}) and detector number of these two detector groups. This may suggest that the GW150914-GBM is likely caused by a sudden increase of the local particles around the spacecraft. 

\floattable
\begin{deluxetable}{ccCcccc}
\tablecaption{Fermi/GBM-Swift jointly-detected GRBs with similar incident angles as the GW150914. \label{tab:grbs}}
\tablecolumns{7}
\tablenum{1}
\tablewidth{0pt}
\tablehead{
\colhead{GRB name} &
\colhead{GBM name} &
\colhead{$\theta$\tablenotemark{a}} &
\colhead{$\phi$\tablenotemark{a}} & \colhead{Ang\_dist\tablenotemark{b}} &
\colhead{GW location\#\tablenotemark{c}} &  \colhead{ratio\tablenotemark{d}}  \\
\colhead{} & \colhead{} & \colhead{(deg)} & \colhead{(deg)} &
\colhead{(deg)} & \colhead{} & \colhead{}
}
\startdata
GRB140716A &	GRB140716436    &     134.1 &     352.2   	&    8.8	&   5/6/7/8  &  1.4 $\pm$ 0.2 \\
GRB140817A &	GRB140817293    &    131.5  &     10.7   	&    9.5	&   7/8   	&   1.7 $\pm$ 0.2 \\
GRB150607A &	GRB150607330    &     130.1 &      7.7  	&     9.3    &   6/7/8  	&  [1.7, 11.2]\tablenotemark{e}  \\
\enddata
\tablenotetext{a}{Incident angle in spacecraft coordinates where zenith corresponds to $\theta$ = 0$\arcdeg$.}
\tablenotetext{b}{The typical angular distance between GRB and LIGO locations of GW150914.}
\tablenotetext{c}{Row number of the GW locations in Table 2 in \citet{connaughton_fermi_2016}.}
\tablenotetext{d}{Ratio of photons detected by good-viewing and bad-viewing detectors (see the text).}
\tablenotetext{e}{90\% confidence interval.}
\end{deluxetable}

\section{Discussion and Summary} \label{sec:dis}

By using a straightforward but sophisticated method, we find that the excess of GW150914-GBM, which is suggested to be a possible counterpart of GW150914 \citep{connaughton_fermi_2016}, is mostly contributed by those GBM detectors that are not expected to see much radiation from the LIGO localization region of GW150914, whereas detectors (NaI 2, NaI 4, NaI 5, BGO 0) with good exposures to the GW event did not see any significant excess beyond the background fluctuation. The distribution of excess signals among detectors of GW150914-GBM is very inconsistent from the expected value that has been demonstrated by GBM-Swift jointly-detected GRBs with similar incident angles as GW150914. However, $\sim$ 2000 GRBs have been successfully localized based on such distribution \citep{kienlin_second_2014}. Therefore, we conclude that the GW150914-GBM is very unlikely associated with GW150914, and the origin of the excess in bad-viewing detectors, which is not the focus of this paper though, is possibly due to some sudden increase in the local particles, while the relatively small excess in the good-viewing detectors is not distinguishable from the background fluctuation. 

Since GW150914-GBM is the only event found by GBM that is possibly associated with GW150914 in a comprehensive search, the rejection of it to be an astrophysical event in the LIGO localizations means that GBM did not detect any significant EM counterpart of GW150914. \citet{liu_exact_2009} predicted that the merging of two astrophysical black holes will not produce any electromagnetic emission, based on the exact global solution of the dynamic and evolving metric for matter falling onto a pre-existing black hole in the frame of a distant external observer. As of today, the non-detection of EM counterpart of GW150914 by a series of telescopes \citep{abbott_localization_2016} including Fermi/GBM is consistent with this prediction \citep{zhang_how_2016}.

However, it is commonly believed that the in-falling matter to a black hole must approach asymptotically to but never cross the event horizon of the black hole in the frame of a distant external observer \citep[e.g.][]{vachaspati_observation_2007, oppenheimer_continued_1939}; in this case the merging of two such objects will produce both GW and strong EM radiations \citep{zhang_how_2016}, through the well-known Blandford-Znajek (BZ) mechanism \citep{blandford_electromagnetic_1977}. Because the relativistic jet produced through the BZ mechanism for a GW150914-like event may not point to the observer, the non-detection of strong EM radiation from just one GW150914-like event does not distinguish between the above two models; therefore, a statistically significant sample of GW150914-like events with comprehensive EM search is highly demanded.

Thanks to the very wide instantaneous FOV, broad energy coverage and moderate localization capability, Fermi/GBM is one of the best telescopes to find the purported EM counterparts of the gravitational wave events. As more joint observations are undergoing between LIGO and Fermi/GBM, it is very promising for Fermi/GBM to either find some EM counterparts or put a strong constraint on the EM radiation of GW150914-like events.

\acknowledgments
I thank Shuang-Nan Zhang, Fangjun Lu and Liming Song for very helpful discussions and encouragement. I am grateful to the Fermi/GBM team and NASA for the public data and software. This work is partially supported by the ''Hundred Talents'' program funded by the Chinese Academy of Sciences (CAS), the National Basic Research Program ("973" Program) of China (Grants 2014CB845800) and the Strategic Priority Research Program on Space Science, the Chinese Academy of Sciences, Grant No. XDA04010202 and XDA04010300.

\end{document}